\documentclass{article}

\usepackage{arxiv}

\usepackage[utf8]{inputenc} 
\usepackage[T1]{fontenc}    
\usepackage{caption}
\usepackage{hyperref}      
\usepackage{booktabs}
\usepackage{amsmath}
\usepackage{subcaption}
\usepackage{pgffor}    
\usepackage{authblk}       
\usepackage{multirow}
\usepackage{amsfonts}                  
\usepackage{graphicx}
\usepackage{adjustbox}
\usepackage[style=authoryear,backend=biber]{biblatex}
\title{A Personalized MOOC Learning Group and Course Recommendation Method Based on Graph Neural Network and Social Network Analysis}

\date{}
\newif\ifuniqueAffiliation

\uniqueAffiliationtrue

\ifuniqueAffiliation
\author{
	\textbf{Zijin Luo}\textsuperscript{1, a*} 
	\textbf{Xu Wang}\textsuperscript{†2, 6, b*} 
	\textbf{Yiquan Wang}\textsuperscript{3, 6, c*}
	\textbf{Haotian Zhang}\textsuperscript{4, d}
	\textbf{Zhuangzhuang Li}\textsuperscript{5, e}
}

\affil{
	\textsuperscript{1}College of Mathematics, Jilin University, Changchun, Jilin, 130000, China \\
	\textsuperscript{2}College of Communication Engineering, Jilin University, Changchun, Jilin, 130000, China \\
	\textsuperscript{3}College of Mathematics and System Science, Xinjiang University, Urumqi, Xinjiang, 830046, China \\
	\textsuperscript{4}College of Computer Science and Technology, Jilin University, Changchun, Jilin, 130000, China \\
	\textsuperscript{5}Department of Electronic Engineering, Tsinghua University, Beijing, 100084 \\
	\textsuperscript{6}Shenzhen X-Institute, Shenzhen, China, 518055 \\
	\vspace{1em}
	\textsuperscript{a}luozj1020@mails.jlu.edu.cn\hspace{2em}
	\textsuperscript{†b}wangxu2020@mails.jlu.edu.cn\hspace{2em}
	\textsuperscript{c}ethan@stu.xju.edu.cn\\
	\textsuperscript{d}zhanght2120@mails.jlu.edu.cn\hspace{2em}
	\textsuperscript{e}lzz22@mails.tsinghua.edu.cn\\
	\vspace{1em}
	\text{*:Zijing Luo, Xu Wang and Yiquan Wang contributed equally to this work. They are co-first authors.}
}
\fi

\renewcommand{\shorttitle}{}

\hypersetup{
pdftitle={A Personalized MOOC Learning Group and Course Recommendation Method Based on Graph Neural Network and Social Network Analysis},
pdfsubject={cs},
pdfauthor={Zijin Luo1, Xu Wang, Yiquan Wang, Haotian Zhang, Zhuangzhuang Li},
pdfkeywords={Massive open online courses; Social network analysis; Graph neural networks; Personalized recommendations; AI assisted teaching},
}

\begin{document}
\maketitle

\begin{abstract}
In order to enhance students' initiative and participation in MOOC learning, this study constructed a multi-level network model based on Social Network Analysis (SNA). The model makes use of data pertaining to nearly 40,000 users and tens of thousands of courses from various higher education MOOC platforms. Furthermore, an AI-based assistant has been developed which utilises the collected data to provide personalised recommendations regarding courses and study groups for students. The objective is to examine the relationship between students' course selection preferences and their academic interest levels. Based on the results of the relationship analysis, the AI assistant employs technologies such as GNN to recommend suitable courses and study groups to students. This study offers new insights into the potential of personalised teaching on MOOC platforms, demonstrating the value of data-driven and AI-assisted methods in improving the quality of online learning experiences, increasing student engagement, and enhancing learning outcomes.
\end{abstract}

	
	\section{Introduction}
	The rapid popularisation of online learning has led to a significant increase in the use of Massive Open Online Courses (MOOC) as a mode of global education. This is due to the fact that MOOC are open to all and can be scaled up to accommodate large numbers of students. In particular, the application of MOOC in China has increased significantly during the period of the global pandemic caused by the COVID-2019. Despite the extensive resources provided by MOOC platforms for learners, many students still encounter difficulties such as low engagement and insufficient learning motivation during the learning process(\cite{grunspan2014, ren2019, wong2019, vilkova2021}). This raises the question of how to effectively enhance students' learning experience and increase their sense of participation in online education.
	
\begin{figure}[h]
		\centering
		\includegraphics[width=0.8\textwidth]{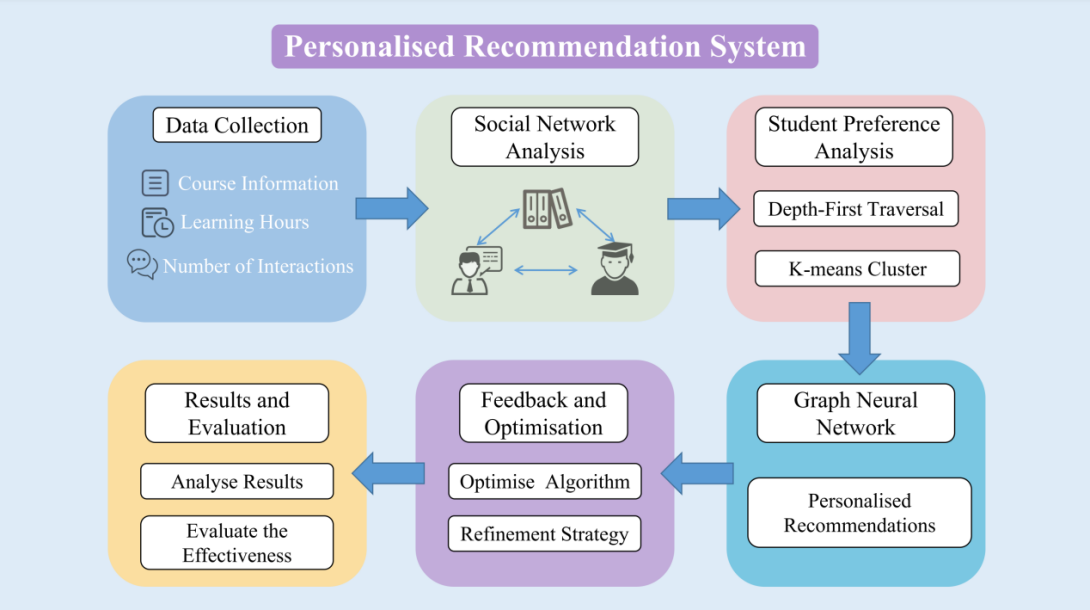}
		\caption{Flow Chart of Personalized Recommendation System}
		\label{fig:flowchart}
	\end{figure}

	The existing research evidence indicates that social interaction among students has a significant impact on learning outcomes(\cite{sunar2016, almatrafi2018, wise2018}). The application of Social Network Analysis (SNA) can facilitate a more profound comprehension of students' interactive patterns in learning, thereby enabling the optimisation of their learning pathways(\cite{grunspan2014, claros2015, cela2015}). However, existing research has concentrated on teacher-student interaction in traditional classrooms, with less attention paid to students' course selection behaviour and its impact on learning initiative and participation on MOOC platforms(\cite{wang2019}). To address this gap, this study has constructed a multi-level model through Social Network Analysis to analyse the relationship between students' course preferences and learning behaviour. Based on this analysis, a personalised recommendation system has been proposed, which is based on Graph Neural Network (GNN). The schematic diagram of the system's construction is shown in Figure \ref{fig:flowchart}.
	
	\section{Related Works}
	
	\subsection{Social Network Analysis (SNA)}
	The field of education has widely embraced the application of Social Network Analysis (SNA), particularly in the examination of students' interactive behaviour and collaborative learning patterns. Kizilcec et al. employed cluster analysis to categorise learners according to their learning patterns and engagement, thereby elucidating the learning behaviours of disparate student groups(\cite{kizilcec2013}). The aforementioned studies demonstrate that SNA is a valuable tool for elucidating the relationship between students' social interactions and learning outcomes. By analysing the interactive network between learners, SNA enables educators to identify learners' participation patterns and potential issues, thereby providing a foundation for targeted interventions. In particular, on MOOC platforms, SNA is employed to examine students' interactive characteristics, anticipate potential dropout risks, and refine the formation of study groups. By leveraging this approach, SNA furnishes data-driven insights to inform personalised learning recommendations, enabling students to access learning resources and collaboration opportunities that align more closely with their needs, thus enhancing their learning experience and effectiveness(\cite{mansur2013, kovanovic2019, wang2022}).
	
	\subsection{Graph Neural Network (GNN)}
	In recent years, there has been a notable advancement in the utilisation of Graph Neural Network (GNN) in the field of education, particularly in the domains of personalised learning recommendations and student modelling. By capturing the intricate interconnections between learners and learning resources, GNN is capable of devising bespoke course recommendations and learning pathways for students. Such models as Graph Convolutional Networks (GCN) (\cite{zhang2019, wang2020, xie2020}) and Graph Attention Networks (GAT) (\cite{velickovic2017}) not only optimise recommendation systems but also enhance students' matching and interest in learning resources. The application of GNN extends the capabilities of traditional educational recommendation systems, providing accurate personalised recommendations based on students' learning history, interactive behaviour, and social network structure. Furthermore, GNN have demonstrated considerable potential in the prediction of student performance (\cite{li2022}), the tracking of knowledge (\cite{ni2023}), and other areas, offering robust support tools for personalised teaching through the analysis of students' learning progress.
	
	\subsection{Group Recommendation Systems in Collaborative Learning}
	The objective of the group recommendation system is to facilitate collaborative learning by providing learning resources for learning groups. This is achieved by analysing students' interaction patterns and learning needs, matching peers with similar learning goals and styles, and optimising collaboration effectiveness. In recent years, the application of similarity analysis-based models (such as PGR-ELM (\cite{liu2017}) and SACML (\cite{wang2019})) and graph convolutional network technology (\cite{wei2019}) has enabled these systems to accurately capture student associations, improve recommendation accuracy, and enhance group collaboration effectiveness (\cite{umer2023)}. The application of data mining technology facilitates the optimisation of the construction and resource allocation of learning groups, ensuring diversity and collaborative potential within the group. This enhances interaction and learning efficiency, and facilitates improved academic performance in both personalised and collaborative learning environments(\cite{zain2014}).
	
	\subsection{AI assisted teaching}
	In recent years, there has been a notable advancement in the utilisation of artificial intelligence (AI) technology in the domain of education(\cite{vargas2023}), particularly in the realms of personalised teaching and learning analytics. The utilisation of AI-assisted teaching employs technologies such as machine learning and natural language processing to provide students with precise learning advice and support(\cite{alqahtani2023}). By analysing a substantial corpus of learner behaviour data, AI systems are able to monitor students' learning progress in real time, identify their interests and needs, and provide personalised course recommendations and learning path planning(\cite{kabudi2021}). Furthermore, AI technology can facilitate collaborative learning, assist students in locating suitable learning partners and groups, and enhance learning motivation and effectiveness(\cite{magnisalis2011}). These applications have expanded the boundaries of traditional education and provided new possibilities for the construction of intelligent and personalised learning environments.
	
	\section{Materials and Methods}
	
	\subsection{Data Collection}
	The data for this study was sourced from a specific Massive Open Online Course (MOOC) platform for higher education, encompassing 37,457 user records and 12,779 courses. The data set includes user identification (ID), occupation, course selection information, learning duration, and interaction frequency. In order to gain insight into the behavioural patterns of learners, this study focused on a sample of 8419 users who had completed 27 or more elective courses. The data provide a wealth of information on user learning behaviour, and also facilitate the analysis of the interactive relationship graph between "course student teacher" based on multidimensional social networks. Students who select a greater number of courses demonstrate a higher level of engagement in their learning activities. The reputation and teaching methods of the instructors have a direct impact on the courses that students choose to take. Courses with high levels of interconnectivity are typically taught by experienced instructors, thereby providing a foundation for the study of learners' learning habits, interaction patterns, and course selection preferences on MOOC platforms.
	
	\begin{table}[h]
		\centering
		\caption{Course Information Statistics}
		\label{tab:course_info}
		\begin{tabular}{@{}lccclcc@{}}
			\toprule
			Course & Number & Course & Number & Course & Number \\
			\midrule
			Electronics & 12 & Arts \& Culture & 10 & Computer & 793 \\
			Civil Engineer & 11 & Histories & 13 & Economics & 553 \\
			Medicine & 15 & Bio \& Life Sci & 8 & Science & 5 \\
			Math & 5 & Communication & 2 & Psychology & 119 \\
			Management & 947 & Science & 1286 & Philosophy & 105 \\
			Engineering & 2476 & Law & 446 & Test & 15 \\
			National Boutique & 1316 & Art \& Design & 603 & Medicine & 825 \\
			Foreign Languages & 569 & Literature & 484 & Education & 603 \\
			Agriculture & 262 & General Studies & 105 & Foundation & 17 \\
			\midrule
			\multicolumn{2}{c}{Total Course Number} & 27 & \multicolumn{2}{c}{Total User Records} & 11605 \\
			\bottomrule
		\end{tabular}
	\end{table}
	
	\subsection{Analysis Tools and Models}
	This study has developed an artificial intelligence assistant that employs Social network analysis (SNA) and Graph neural network (GNN) techniques to elucidate the interaction patterns between students, courses, teachers, and classmates during the course selection process. This has the objective of promoting an understanding of their social network structure. Furthermore, the AI assistant is capable of discerning the complex interconnections between students, courses, and educators through GNN, thus enhancing the precision of personalised recommendations. This multi-level analytical approach not only provides a comprehensive understanding of learners' behaviours but also establishes the foundation for developing personalised recommendation systems, which assist students in making more informed decisions regarding course selection and study group arrangements, thereby improving their learning experience.
	
	\subsection{User Preference Analysis}
	The AI assistant employed a deep first traversal algorithm to analyse the connection relationships between students and entities such as courses, schools and teachers. This analysis revealed differences in course selection and learning methods among different student groups. The findings of this analysis provide valuable insights that can inform the development of personalised recommendation systems and instructional design, ensuring that courses are better aligned with students' preferences and needs.
	
	\subsection{The Relationship Between Preference and Learning Engagement}
	The AI-powered personalized recommendation system employs K-means clustering analysis to examine students' preferences regarding course selection. Additionally, it assesses students' engagement in the learning process through indicators such as learning duration, interaction frequency, and feedback frequency. To quantify the correlation between course selection preferences and learning engagement, the study utilized Rand and Pearson correlation coefficients. This approach aimed to elucidate the influence of course selection preferences on learning outcomes and to provide support for the implementation of personalized recommendation systems to enhance academic performance.
	
	\subsection{Analysis of Learning Groups Based on GNN}
	
	\subsubsection{Building a Study Group}
	Based on the course preferences expressed by students, artificial intelligence (AI) assistants are employed to group them into study groups. This grouping method allows students to motivate each other through collaboration and communication in group learning, thereby improving their learning motivation and academic performance, and promoting a deeper learning experience.
	
	\subsubsection{Application of GNN Model}
\begin{figure}[h]
		\centering
		\includegraphics[width=0.7\textwidth]{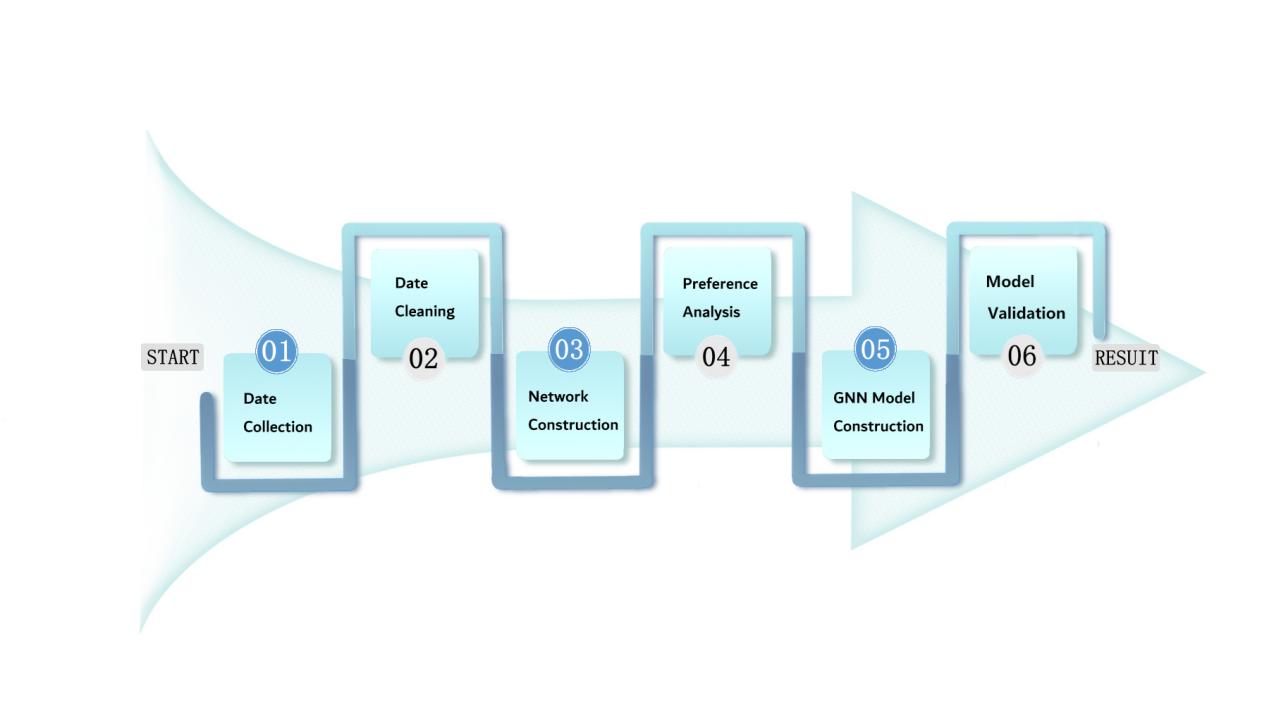}
		\caption{The GNN flowchart}
		\label{fig:gnn_flowchart}
	\end{figure}

\begin{figure}[h]
		\centering
		\includegraphics[width=0.7\textwidth]{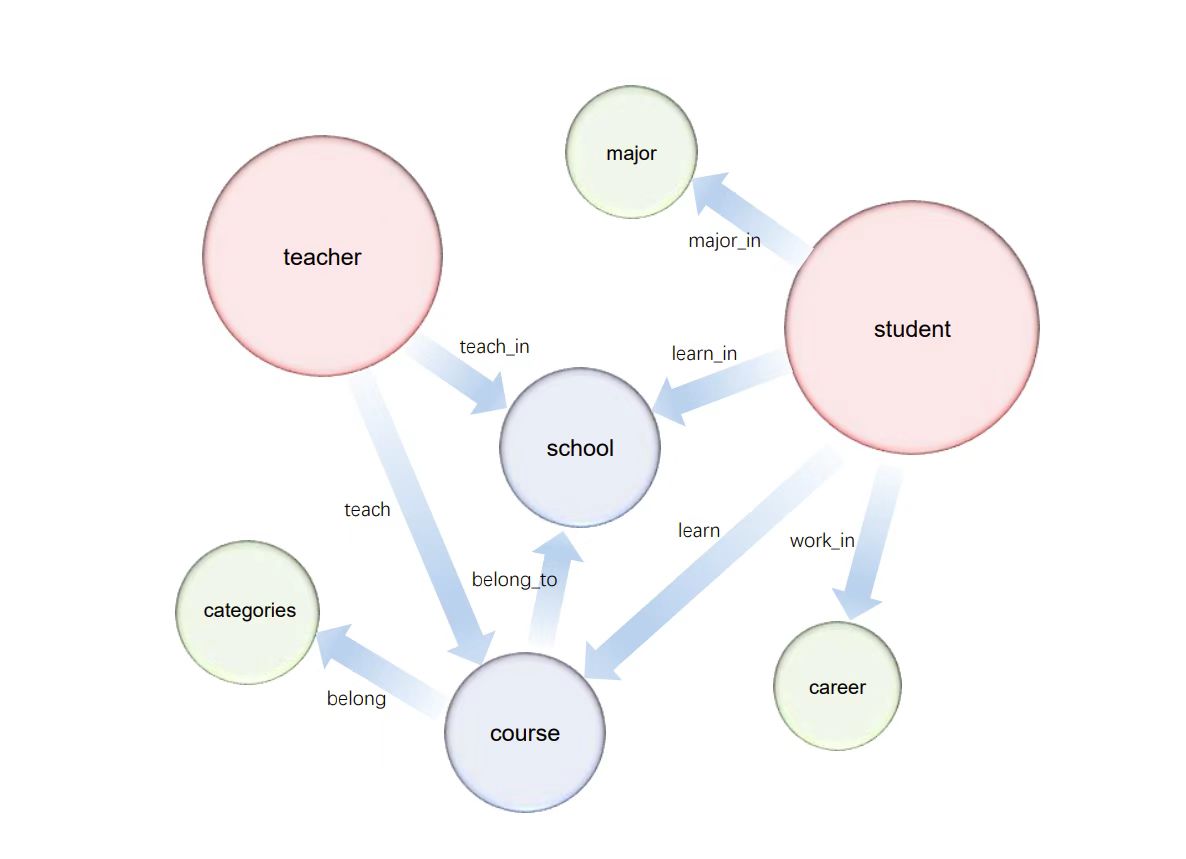}
		\caption{The Simplified GNN Model: Student-Course-Teacher Graph}
		\label{fig:gnn_model}
	\end{figure}

	In order to gain a deeper insight into the collaborative efficacy of learning groups, this study employed the Graph Neural Network (GNN) model for the purpose of analysing the interactive relationships that exist within the group. The GNN flowchart is presented in Figure \ref{fig:gnn_flowchart}. A social network structure of "course student teacher" has been constructed, as illustrated in Figure \ref{fig:gnn_model}, with the objective of quantifying the collaborative effectiveness among group members, analysing the performance differences of students in the same learning group, capturing the interaction patterns of members in the learning group and their impact on learning outcomes, and predicting their learning outcomes. By means of analysis, AI teaching assistants are able to identify optimal learning groups and furnish data in support of subsequent optimisation, thereby enhancing the efficiency of collaborative learning, enriching students' learning experience, and conferring overall benefits.
	
	The GNN flowchart illustrates the comprehensive research process, encompassing data collection, data cleansing, social network construction, preference analysis, and model validation. This graphical representation facilitates a clear understanding of the research methodology.
	
	The simplified GNN model, which depicts the relationship between students, courses, and teachers, demonstrates how GNN captures the association between students and learning resources through graph structure.
	
	\section{Results}
	
	\subsection{Establishment and Evaluation of Social Networks}
	This study analysed the learning data of MOOC platforms and constructed a knowledge graph containing students, teachers, courses, and their interrelationships using the Neo4j graph database. This graph provides an invaluable tool for a deeper understanding of students' behavioural patterns in course selection, learning interaction, and teacher support. It reveals students' behavioural patterns in course selection, learning interaction, and teacher support, allowing for the design of more personalised recommendation systems.
	
	\subsubsection{Node Level Indicators}
	At the node level, the following key indicators were subjected to analysis:
	\begin{enumerate}
		\item Number of entities and relationships: The number of entities and relationships in the graph can be counted in order to evaluate the diversity of the learning resources available on the MOOC platform. This analysis illustrates how the platform offers students the opportunity to pursue personalised learning pathways through a diverse range of learning resources and rich interactive relationships.
		\item Node Degree: The degree of each node is indicative of its importance within the network, with a higher degree reflecting a greater impact on learning behaviour or course recommendations. For example, courses with higher degrees are typically core courses with high student engagement and substantial learning outcomes.
		\item Hierarchical structure: The hierarchical structure of the knowledge graph illustrates how students gradually construct their learning paths with the guidance of teachers and courses. This structure enables educational platforms to gain a deeper understanding of students' learning processes and provide course recommendations that are more aligned with individual learning needs.
	\end{enumerate}
	
	Tables \ref{tab:entity_nodes} and \ref{tab:entity_relationships}, respectively, demonstrate the richness of the data set and the completeness of the graph by showing the number of nodes and relationships between different entity types. They intuitively reflect the complex relationships between learners, courses, and teachers on the MOOC platform, and provide support for personalised recommendation systems, enabling them to optimise course recommendations based on students' specific needs.
	
	\begin{table}[h]
		\centering
		\caption{Number of entity nodes per category, fields included and related descriptions}
		\label{tab:entity_nodes}
		\begin{tabular}{@{}llllc@{}}
			\toprule
			Node type & Node field name & Field Description & Number of nodes & Total number of nodes \\
			\midrule
			\multirow{6}{*}{\centering Student} & name & User's name & \multirow{6}{*}{\centering 6363} & \multirow{17}{*}{\centering 72211} \\
			& id & User ID & & \\
			& url & User's personal page & & \\
			& learning\_time & Total user learning hours & & \\
			& response & Total number of user responses & & \\
			& likes & Total number of user likes & & \\
			\cmidrule(lr){1-4}
			\multirow{4}{*}{\centering Course} & name & Course Name & \multirow{4}{*}{\centering 24703} & \\
			& id & Course ID & & \\
			& url & Course Page Website & & \\
			& num & Number of electors & & \\
			\cmidrule(lr){1-4}
			\multirow{3}{*}{\centering Teacher} & name & Teacher's name & \multirow{3}{*}{\centering 39554} & \\
			& id & Teacher ID & & \\
			& career & Title of Teachers & & \\
			\cmidrule(lr){1-4}
			School & name & Name of the school & 1196 & \\
			\cmidrule(lr){1-4}
			Career & name & Occupation of users & 3 & \\
			\cmidrule(lr){1-4}
			Major & name & User's specialty & 360 & \\
			\cmidrule(lr){1-4}
			Categories & name & Classification of Courses & 32 & \\
			\bottomrule
		\end{tabular}
	\end{table}

\begin{table}[h]
	\centering
	\caption{Number of Entity Nodes per Class, Fields Contained, and Related Descriptions}
	\label{tab:entity_relationships}
	\begin{tabular}{@{}llllc@{}}
		\toprule
		Relationship name & Arc-head node types & Arc-tail node type & Number of relationships & Total number of relationships \\
		\midrule
		Belong & Course & Categories & 11704 & \multirow{8}{*}{\centering 704785} \\
		Belong\_to & Course & School & 24703 & \\
		Learn & Student & Course & 519237 & \\
		Learn\_in & Student & School & 744 & \\
		Major\_in & Student & Major & 2097 & \\
		Teach & Teacher & Course & 100382 & \\
		Teach\_in & Teacher & School & 39555 & \\
		Work\_in & Student & Career & 6363 & \\
		\bottomrule
	\end{tabular}
\end{table}
	
	The aforementioned node types serve to illustrate the significant participants and entities that are integral to the MOOC ecosystem.
	
	\subsection{Analysis Results of Preferences and Learning Initiative}
	In order to gain a deeper insight into the relationship between students' course preferences and their learning engagement, we employed the K-means clustering algorithm to categorise students' course selection preferences. This was followed by the verification of the significant relationship between these preferences and learning initiative through the use of the Rand similarity coefficient and the chi-square test.
	
	\subsubsection{Cluster Results}
\begin{figure}[h]
		\centering
		\begin{subfigure}[b]{0.33\textwidth}
			\centering
			\includegraphics[width=\textwidth]{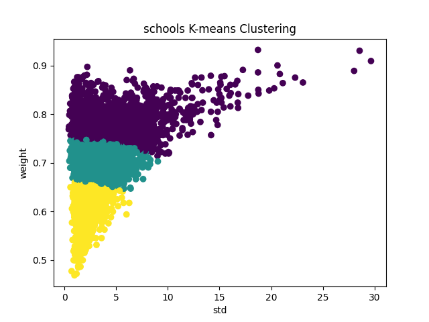}
			\caption{}
			\label{fig:image1}
		\end{subfigure}
		\hfill
		\begin{subfigure}[b]{0.33\textwidth}
			\centering
			\includegraphics[width=\textwidth]{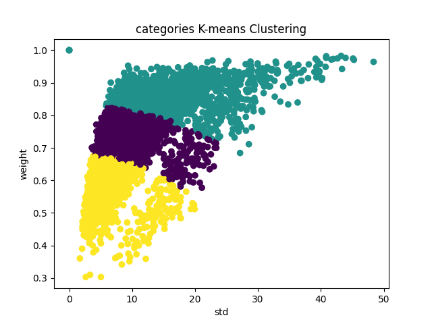}
			\caption{}
			\label{fig:image2}
		\end{subfigure}
		\hfill
		\begin{subfigure}[b]{0.33\textwidth}
			\centering
			\includegraphics[width=\textwidth]{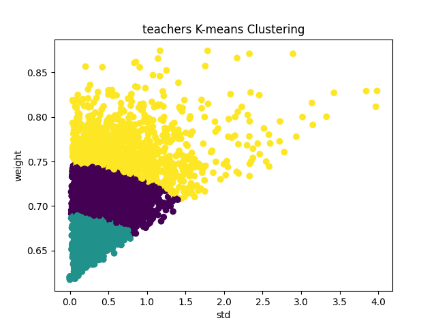}
			\caption{}
			\label{fig:image3}
		\end{subfigure}
		\caption{Clustering results of different preferences}
		\label{fig:clustering_results}
	\end{figure}

	Figure \ref{fig:clustering_results} illustrates the clustering results of different course preference groups. By analysing the course preferences at the high, medium, and low levels, we found a close positive correlation between course selection preferences and learning engagement. Notably, students from high preference groups exhibited higher levels of participation in the learning process, interacting more frequently with course content and actively participating in group discussions. This indicates that personalised recommendation systems effectively enhance students' learning initiative and performance by matching their preferences and needs.
	
	The images are presented in a sequential order, beginning with the school, followed by the category and concluding with the teacher preferences. The correlation between the three clusters, delineated from the lowest to the highest point on the vertical axis, increases in sequence. The horizontal axis represents the course variance of each category, while the vertical axis represents the proportion of courses selected from the top-ranked categories to the total number of courses.
	
	\subsubsection{Chi-square Test Results}
	In order to more accurately reflect students' learning initiative, this study selected the following three indicators: total learning time, average learning time per class, and number of likes received. However, there is an overlap between the total learning time and the average learning time, and thus they should not be used simultaneously. Furthermore, although the number of likes can, to some extent, reflect learners' feedback on the content, it is more influenced by other learners' attention to specific issues or content, and therefore cannot be used as a direct measure of an individual's learning initiative. In light of the above considerations, for the case of using different types of indicators, we have considered the indicator combinations and labels shown in Table \ref{tab:notation_scenarios}.
	
	\begin{table}[h]
		\centering
		\caption{Notation for different scenarios}
		\label{tab:notation_scenarios}
		\begin{tabular}{@{}lcc@{}}
			\toprule
			& Adoption of the number of likes & Not using the number of likes \\
			\midrule
			Adoption of total learning hours & FF & FT \\
			Adoption of average learning hours & TF & TT \\
			\bottomrule
		\end{tabular}
	\end{table}
	
	Furthermore, the notable correlation between course selection preference and learning engagement was substantiated through a chi-square test, with all p-values less than 0.05, indicating a statistically significant correlation. Personalised recommendation systems can facilitate the provision of tailored learning pathways based on students' course preferences and learning behaviours, thereby enhancing students' learning motivation.
	
	To illustrate, the significance level of 0.05 was employed to calculate the p-values for each scenario as follows:
	
	\begin{table}[h]
		\centering
		\caption{p-values for the four scenarios and three preferences described above}
		\label{tab:p_values}
		\begin{tabular}{@{}lcccc@{}}
			\toprule
			& FF & FT & TF & TT \\
			\midrule
			School Preference & $2.29 \times 10^{-19}$ & $1.69 \times 10^{-19}$ & $1.61 \times 10^{-3}$ & $1.62 \times 10^{-5}$ \\
			Type Preference & $6.27 \times 10^{-2}$ & $2.67 \times 10^{-2}$ & 0.18 & 0.12 \\
			Teacher Preference & $9.79 \times 10^{-2}$ & $2.68 \times 10^{-2}$ & $3.81 \times 10^{-3}$ & $9.53 \times 10^{-4}$ \\
			\bottomrule
		\end{tabular}
	\end{table}
	
	\subsubsection{Comparison of Different Occupational Groups}
\begin{figure}[h]
		\centering
		\begin{subfigure}[b]{0.33\textwidth}
			\centering
			\includegraphics[width=\textwidth]{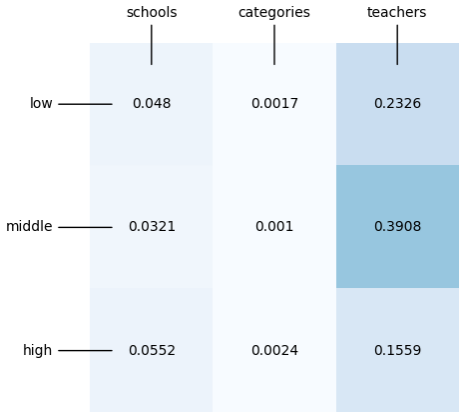}
			\caption{rand index students}
			\label{fig:rand_index_students}
		\end{subfigure}
		\hfill
		\begin{subfigure}[b]{0.33\textwidth}
			\centering
			\includegraphics[width=\textwidth]{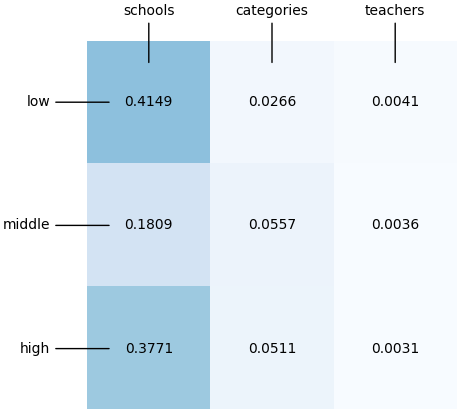}
			\caption{rand index professionals}
			\label{fig:rand_index_professionals}
		\end{subfigure}
		\hfill
		\begin{subfigure}[b]{0.33\textwidth}
			\centering
			\includegraphics[width=\textwidth]{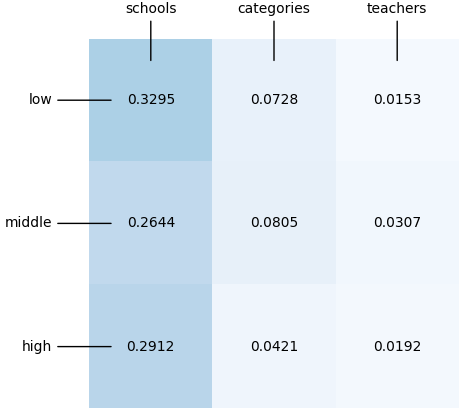}
			\caption{rand index others}
			\label{fig:rand_index_others}
		\end{subfigure}
		\caption{Rand Index for students, working professionals, and other groups}
		\label{fig:rand_index}
	\end{figure}

	Furthermore, the learning behaviours of students, working professionals and other groups were analysed, and it was found that an individual's occupational background has a significant impact on their course selection preferences and learning initiative. Figure \ref{fig:rand_index} illustrates the relationship between course preferences and learning engagement among these three groups. Previous research has demonstrated that working professionals tend to choose courses related to career development and have a higher level of initiative in learning, while students are more interested and have a lower level of initiative. Therefore, it can be concluded that an individual's professional background and motivation should be given importance in the context of personalised recommendations.
	
	\section{Comparative Analysis}
	In order to evaluate the effectiveness of AI assisted teaching systems and their acceptance among different user groups, we conducted a questionnaire survey and collected comparative data before and after system use. A total of 200 valid questionnaires were obtained. The results indicate that the system has significant effectiveness and huge market potential. Specifically, 75\% of respondents are aware that course recommendations on MOOC platforms are driven by AI technology; 73\% of respondents trust course content recommended by AI; 66\% of respondents expressed satisfaction with the overall experience of AI course recommendations. In addition, the survey shows that young users have a higher acceptance of the system, especially among users who have already used the system, and generally give positive feedback, indicating that the young group has a higher recognition and willingness to use the system.
	
\begin{figure}[h]
		\centering
		\includegraphics[width=0.8\textwidth]{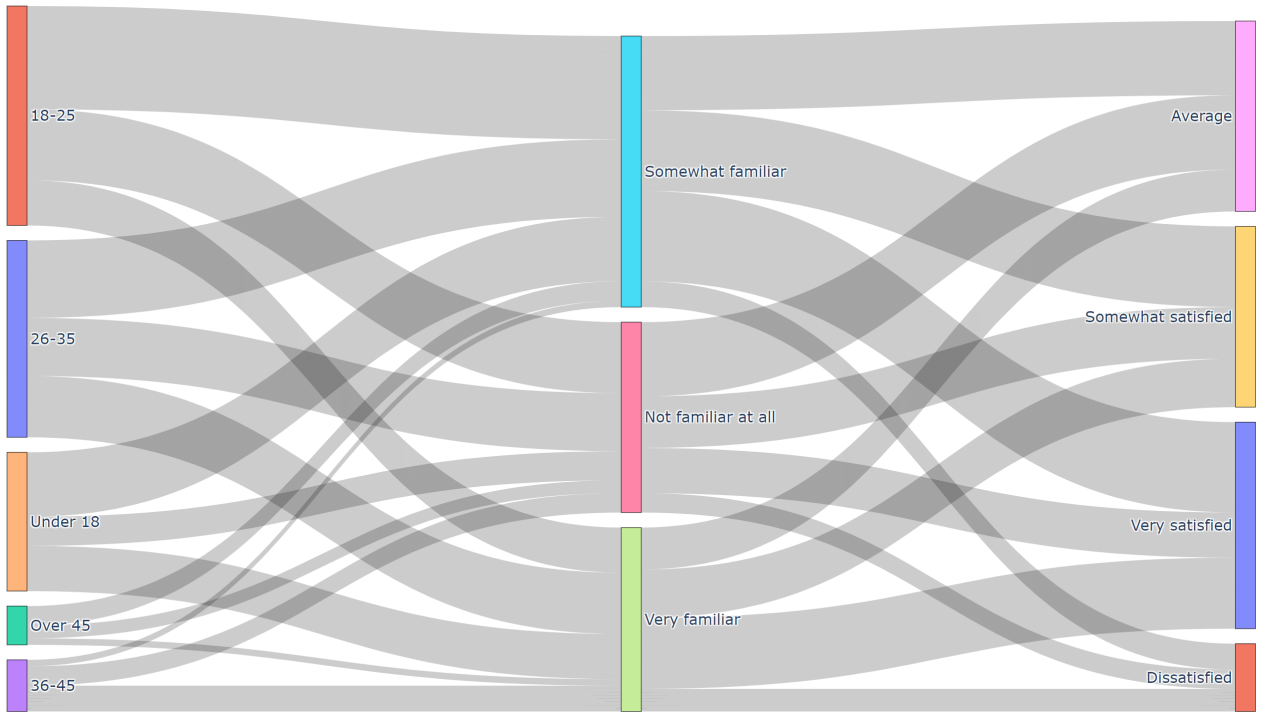}
		\caption{Sankey Diagram of Age, AI Knowledge Level, and Group Satisfaction}
		\label{fig:sankey_diagram}
	\end{figure}

	Figure \ref{fig:sankey_diagram} is a Sankey plot of age, artificial intelligence knowledge level, and group satisfaction. The detailed information of the questionnaire can be found in the supplementary materials.
	
	\section{Conclusion}
	This study combines the development of an AI teaching assistant and uses social network analysis (SNA) and knowledge graph technology to deeply investigate the correlation between course preferences and learning engagement of MOOC learners in higher education. The study employed the Rand similarity coefficient, Pearson correlation, and chi-square independence test to verify the significant impact of course preference on learning engagement. In contrast to previous studies, which have primarily concentrated on the interactions between learners and educators or peers, this study broadens the analytical scope to encompass a range of entities, including courses, institutions, and course types. This approach offers a novel perspective for comprehending the intricacies of the MOOC learning ecosystem.
	
	In conclusion, this study enhances our comprehension of MOOC online learning behaviour and furnishes innovative personalised learning solutions. These findings provide valuable insights for future online education practices, enhancing students' learning experiences and advancing the design of personalized learning paths on educational platforms. Personalized recommendation systems enable online education to more accurately meet the needs of learners from diverse backgrounds, fostering a more autonomous, collaborative, and efficient learning experience. These insights are poised to profoundly impact the field of education and are expected to be further promoted and applied in future educational practices.
	
	\section{Discussion}
	
	\subsection{Educational Significance of Research Results}
	This study combines AI-assisted teaching, social network analysis, and knowledge graphs to reveal a significant link between MOOC learners' course preferences and their engagement. Students with higher preferences engage more, highlighting the importance of considering learners' interests in designing personalized courses. Recommending courses based on preferences enhances outcomes, and AI-powered recommendation systems using graph neural networks can boost engagement and foster collaborative learning, supporting personalized learning paths.
	
	\subsection{Application Suggestions in Educational Practice}
	Future educational platforms should use tools like social network analysis and graph convolutional networks to develop intelligent, flexible AI-based personalized learning support systems. Educators can help students build learning paths aligned with their needs, enhancing participation. Platforms should optimize learning group formation through recommendation systems to improve collaborative learning efficiency, enriching personal experiences and the learning community, and advancing online education.
	
	\subsection{Future Research Directions}
	Further research should deepen understanding of dynamic social network analysis and how learner behavior evolves over time to grasp long-term learning patterns. Enhancing AI recommendation systems by incorporating factors like learning styles and personality traits can improve effectiveness, ensuring each learner accesses the most suitable path. These advancements provide strong support for implementing personalized learning in future educational practices.
	
	\section*{Data availability}
	Data will be made available on request.
	
\printbibliography
\clearpage
\section*{Supplementary Materials}
\appendix
\setcounter{section}{0}
\renewcommand{\thesection}{\arabic{section}} 
\section{Design of Questionnaire Survey}

\subsection{Introduction}
In order to assess the effectiveness of AI-assisted teaching systems and their acceptance by different user groups, we conducted a questionnaire survey and collected comparative data before and after system use.

\subsection{Demographic Information}
In this section, by collecting basic background information of students, we can help analyze the acceptance and experience differences of the AI-assisted teaching system among different background groups (gender, age, grade, major, occupation, etc.).

\begin{enumerate}
	\item Gender
	\begin{itemize}
		\item Male
		\item Female
		\item Other
	\end{itemize}
	
	\item Age
	\begin{itemize}
		\item Under 18
		\item 18-25
		\item 26-35
		\item 36-45
		\item Over 45
	\end{itemize}
	
	\item Grade/Occupation
	\begin{itemize}
		\item Undergraduate
		\item Master's student
		\item Doctoral student
		\item Working professional
		\item Other (please specify)
	\end{itemize}
	
	\item Major
	\begin{itemize}
		\item Science and Engineering
		\item Social Sciences
		\item Literature and Humanities
		\item Arts and Design
		\item Other (please specify)
	\end{itemize}
	
	\item Do you have a full-time or part-time job?
	\begin{itemize}
		\item Yes, full-time
		\item Yes, part-time
		\item No
	\end{itemize}
\end{enumerate}

\subsection{Awareness and Trust in AI-Assisted Teaching}

\begin{enumerate}
	\item AI Awareness
	\begin{itemize}
		\item Are you aware that the course recommendations on MOOC platforms are supported by AI systems?
		\begin{itemize}
			\item Yes
			\item No
			\item Heard of it but do not understand the specific functions
		\end{itemize}
	\end{itemize}
	
	\item AI Trust
	\begin{itemize}
		\item Do you think the courses recommended by AI match your learning interests and goals?
		\item To what extent do you rely on AI recommendations when choosing courses?
		\item Do you believe AI technology can better understand your learning needs?
	\end{itemize}
\end{enumerate}

\subsection{Experience of Using Personalized Recommendation Systems}
\begin{itemize}
	\item How satisfied are you with the overall course recommendation system by AI?
	\item Have the courses recommended by the AI system helped you better achieve your learning goals?
	\item Do you think the courses recommended by AI fit your learning schedule?
\end{itemize}

\subsection{Suggestions for AI-Assisted Teaching Systems}

\begin{enumerate}
	\item Technical Experience
	\begin{itemize}
		\item Have you encountered technical issues while using the AI recommendation system?
		\item How is the experience of using the AI system on different devices?
	\end{itemize}
	
	\item System Improvement Suggestions
	\begin{itemize}
		\item Personalization of course recommendations
		\item Accuracy of study group matching
		\item System response speed
		\item More flexible cross-platform adaptation
		\item Provide more explainable recommendation bases
		\item Other (please specify)
	\end{itemize}
	
	\item Future Development
	\begin{itemize}
		\item Do you think AI-assisted teaching systems can replace part of traditional teaching?
		\item What features would you like the AI system to add in the future?
	\end{itemize}
\end{enumerate}

\section{Data Analysis}

\subsection{Sample Demographic Characteristics}

\begin{itemize}
	\item Gender Distribution
	\begin{itemize}
		\item Male: 57\% (114 people)
		\item Female: 43\% (86 people)
	\end{itemize}
	
	\item Age Distribution
	\begin{itemize}
		\item Under 18: 19\% (38 people)
		\item 18-25: 35\% (70 people)
		\item 26-35: 31\% (62 people)
		\item 36-45: 8\% (16 people)
		\item Over 45: 7\% (14 people)
	\end{itemize}
	
	\item Grade/Occupational Status
	\begin{itemize}
		\item Undergraduate: 32\% (64 people)
		\item Master's student: 34\% (68 people)
		\item Doctoral student: 14\% (28 people)
		\item Working professional: 18\% (36 people)
		\item Other: 2\% (4 people)
	\end{itemize}
	
	\item Major
	\begin{itemize}
		\item Science and Engineering: 38\% (76 people)
		\item Social Sciences: 38\% (76 people)
		\item Literature and Humanities: 12\% (24 people)
		\item Arts and Design: 8\% (16 people)
		\item Other: 4\% (8 people)
	\end{itemize}
	
	\item Employment Status
	\begin{itemize}
		\item Yes, full-time: 33\% (66 people)
		\item Yes, part-time: 16\% (32 people)
		\item No: 51\% (102 people)
	\end{itemize}
\end{itemize}

Gender and Age: The survey shows that male users account for 57\%, and female users account for 43\%. This indicates higher male participation in the use of AI-assisted teaching systems. Further analysis of gender differences in course selection and learning preferences can help the system better meet the needs of users of different genders.

Grade and Occupational Status: The proportion of undergraduates and master's students is relatively high, accounting for 32\% and 34\%, respectively. This may indicate that AI-assisted teaching systems are more widely used in higher education. It is worth considering designing specific course recommendation strategies for users of different grades to improve learning outcomes.

Major Background: Users from science and engineering and social sciences each account for 38\%, indicating a high demand for technical and social science courses. By analyzing the learning habits of users with different major backgrounds, the course recommendation system can be further optimized.

\subsection{Awareness of AI-Assisted Teaching}

\begin{enumerate}
	\item AI Awareness
	\begin{itemize}
		\item Aware that MOOC platform course recommendations are supported by AI systems: 56\% (112 people)
		\item No: 25\% (50 people)
		\item Heard of it but do not understand specific functions: 19\% (38 people)
	\end{itemize}
	
	\item Understanding of AI Technology in Teaching
	\begin{itemize}
		\item Very familiar: 32\% (64 people)
		\item Somewhat familiar: 38\% (76 people)
		\item Completely unfamiliar: 30\% (60 people)
	\end{itemize}
\end{enumerate}

\subsection{Trust in AI}

\begin{enumerate}
	\item Course Content Matching
	\begin{itemize}
		\item Very matching: 32\% (64 people)
		\item Quite matching: 41\% (82 people)
		\item Average: 18\% (36 people)
		\item Not matching: 9\% (18 people)
	\end{itemize}
	
	\item Dependence on AI Recommendations
	\begin{itemize}
		\item Very dependent: 35\% (70 people)
		\item Generally dependent: 43\% (86 people)
		\item Not dependent: 22\% (44 people)
	\end{itemize}
	
	\item Trust in AI Understanding Learning Needs
	\begin{itemize}
		\item Very trusting: 42\% (84 people)
		\item Quite trusting: 31\% (62 people)
		\item Uncertain: 22\% (44 people)
		\item Not trusting: 5\% (10 people)
	\end{itemize}
\end{enumerate}

Awareness: 56\% of respondents are aware that MOOC platform course recommendations are supported by AI, but 25\% are not. This indicates a need to strengthen the promotion and education of AI technology, especially in higher education institutions.

Trust: 73\% of respondents (the sum of very trusting and quite trusting) have confidence in AI's ability to understand learning needs, showing a high level of trust in AI systems. However, 22\% are uncertain about AI's understanding capabilities, suggesting that user education and transparent information dissemination could enhance trust.

\subsection{User Experience}

\begin{enumerate}
	\item Course Recommendation Experience
	\begin{itemize}
		\item Overall satisfaction:
		\begin{itemize}
			\item Very satisfied: 29\% (58 people)
			\item Quite satisfied: 37\% (74 people)
			\item Average: 19\% (38 people)
			\item Not satisfied: 15\% (30 people)
		\end{itemize}
		
		\item Did AI-recommended courses help achieve learning goals:
		\begin{itemize}
			\item Yes, greatly helped: 33\% (66 people)
			\item Yes, helped somewhat: 43\% (86 people)
			\item No help: 17\% (34 people)
			\item Negative impact: 7\% (14 people)
		\end{itemize}
		
		\item Do courses fit learning schedule:
		\begin{itemize}
			\item Completely fit: 26\% (52 people)
			\item Quite fit: 44\% (88 people)
			\item Average: 22\% (44 people)
			\item Do not fit: 8\% (16 people)
		\end{itemize}
	\end{itemize}
\end{enumerate}

Course Recommendation Experience: 66\% of respondents are satisfied or very satisfied with AI course recommendations, indicating high effectiveness of the AI system in course recommendations. However, 15\% expressed dissatisfaction, which may require further investigation into the reasons, such as course quality or recommendation algorithm accuracy.

Achievement of Learning Goals: 76\% of respondents believe AI recommendations help them achieve learning goals, indicating the potential of AI systems in supporting learning. Collecting more specific feedback on learning goals could help better adjust recommendation algorithms to meet personalized needs.

\subsection{Suggestions for Improvement and Future Development of the AI System}

\begin{enumerate}
	\item Technical Experience
	\begin{itemize}
		\item Frequency of encountering technical issues:
		\begin{itemize}
			\item Frequently: 38\% (76 people)
			\item Occasionally: 40\% (80 people)
			\item Never: 22\% (44 people)
		\end{itemize}
		
		\item Cross-device usage experience:
		\begin{itemize}
			\item Very smooth: 20\% (40 people)
			\item Basically smooth: 41\% (82 people)
			\item Occasionally lagging: 23\% (46 people)
			\item Frequently lagging: 16\% (32 people)
		\end{itemize}
	\end{itemize}
\end{enumerate}

Technical Experience: 78\% of respondents (the sum of frequently and occasionally) have encountered technical issues, indicating a need to improve system stability and user experience. It is recommended to optimize the system to ensure stability under high load conditions and enhance cross-device compatibility and smoothness.

Personalized Recommendations: Enhance the personalization and accuracy of course recommendations to meet the needs of users with different majors, interests, and learning goals. Consider incorporating more user behavior data and preference analysis to enhance the intelligence of the recommendation system.

Study Group Matching: Improve the accuracy of group matching to enhance interaction and engagement among learners. By analyzing users' learning styles, social preferences, and goals, optimize group recommendations to promote collaborative learning.

\section{Conclusion and Recommendations}

Enhance AI Awareness: Strengthen education and promotion for older populations and groups not fully aware of AI technology to improve overall awareness. Consider organizing lectures, workshops, and other forms to enhance users' understanding of AI technology.

Enhance System Stability: Address technical issues, improve system response speed, and cross-platform usage experience. Regular system maintenance and updates are recommended to ensure smooth user experience.

Increase Functionality and Explainability: Develop real-time learning progress feedback features and provide more explanations for recommendation bases to enhance user trust. Consider introducing user feedback mechanisms to involve users in system improvement.

Through the above analysis, we can more comprehensively understand the current status and improvement directions of the AI-assisted teaching system. These recommendations not only help improve user satisfaction and learning outcomes but also provide important references for future research and practice.
\end{document}